\documentclass[aps,showpacs,preprintnumbers,amsmath, amssymb]{revtex4-2}

\usepackage{amsmath}
\usepackage{float}
\usepackage{graphics,epsfig,hyperref}
\usepackage{graphicx}
\usepackage{dcolumn}
\usepackage{bm}
\usepackage{epstopdf}
\usepackage{subfigure}

\def \be {\begin{equation}}
	\def \ee {\end{equation}}
\def \bea {\begin{eqnarray}}
	\def \eea {\end{eqnarray}}

\begin{document}
\baselineskip=0.8 cm
\title{\bf Chaotic motion of particles in the spacetime of a Kerr black hole immersed in swirling universes}
\author{Deshui Cao$^{1}$, Lina Zhang$^{1}$, Songbai Chen$^{1,2}$\footnote{csb3752@hunnu.edu.cn}, Qiyuan Pan$^{1,2}$\footnote{panqiyuan@hunnu.edu.cn}, and Jiliang Jing$^{1,2}$ \footnote{jljing@hunnu.edu.cn}}	
	
\affiliation{$^1$ Department of Physics, Key Laboratory of Low Dimensional Quantum Structures and Quantum Control of Ministry of Education, Institute of Interdisciplinary Studies, and Synergetic Innovation Center for Quantum Effects and Applications, Hunan Normal University,  Changsha, Hunan 410081, China
\\
$^2$ Center for Gravitation and Cosmology, College of Physical Science and Technology, Yangzhou University, Yangzhou 225009, China}

\begin{abstract}
\baselineskip=0.6 cm
\begin{center}
{\bf Abstract}
\end{center}

We investigate the motion of particles in the spacetime of a Kerr black hole immersed in swirling universes. Using the Poincar\'{e} section, fast Lyapunov exponent indicator, bifurcation diagram and basins of attraction, we present the effects of the swirling parameter and the spin parameter on the dynamical behaviors of the motion of particles, and confirm the presence of chaos in the motion of particles in this background spacetime. We find that the swirling parameter can change the range of the spin parameter where the chaos occurs, and vice versa. Moreover, we observe clearly that, regardless of the spin parameter, there exist some self-similar fractal fine structures in the basins boundaries of attractors for the spacetime of a black hole immersed in swirling universes. The combination the swirling parameter and the spin parameter provides richer physics in the motion of particles.

\end{abstract}

\maketitle
\newpage

\section{Introduction}

As an important nonlinear phenomenon, the chaos is a kind of non-periodic motion with high sensitivity to initial conditions. In the chaotic motion, the tiny differences in initial conditions can grow rapidly at exponential rates and so a long-term prediction to the motion is very difficult \cite{Ott,Sprott,Brown,Brown1}. But the chaos is ubiquitous in nature and chaotic systems possess a lot of novel properties not shared by the linear dynamical systems. Thus, there have been accumulated interest to study the chaotic dynamics in various physical fields, including black hole physics \cite{Bombellitf-Calzetta,Letelier-Vieira,Santoprete-Cicogna,LiuLLCQG,Jai-aksonCEL,DaluiMM,Falco-Borrelli1,Falco-Borrelli2}. In general relativity, we can investigate the chaotic motion of particles by considering some spacetimes with complicated geometrical structures or introducing some extra interactions to ensure that the dynamical system of particles is non-integrable. Along this line, Dettmann \emph{et al.} analyzed the phase space for trajectories in multi-black-hole spacetimes, and obtained the fractal basins and chaotic trajectories of the corresponding dynamical system \cite{DettmannFC}. Karas \emph{et al.} studied the chaotic motion of test particles in the spacetime of a black hole immersed in magnetic field \cite{Karas} and Li \emph{et al.} extended the investigation to the chaotic motion of neutral and charged particles in a magnetized Ernst-Schwarzschild spacetime \cite{Ldan}. Moreover, the chaotic phenomena have been investigated in the perturbed Schwarzschild spacetime  \cite{Aguirregabiria,SotaSM,WitzanySS}, the non-standard Kerr black hole spacetime described by Manko-Novikov metric \cite{non-standardKerr1,non-standardKerr2,non-standardKerr3,non-standardKerr4,non-standardKerr5}, the accelerating and rotating black hole spacetime \cite{schen1}, and the disformal rotating black-hole spacetime \cite{ZhouSCPMA}. On the other hand, Varvoglis \emph{et al.} observed the chaotic behavior for the charged particles moving in a magnetic field interacting with gravitational waves \cite{VarvoglisP}. Frolov \emph{et al.} introduced ring strings instead of point particles and found that the ring string dynamics is chaotic even in the Schwarzschild black hole spacetime \cite{FrolovL}. The chaotic behaviors in the ring string dynamics also appear in the Schwarzschild AdS black hole \cite{ZayasT} and Gauss-Bonnet AdS black hole spacetimes \cite{MaWZ}. Recently, Wang \emph{et al.} introduced an extra interaction with the Einstein tensor and studied the chaotic dynamics of a scalar test particle in the Schwarzschild-Melvin black hole spacetime \cite{mschen2}. Additionally, the researchers investigated the chaotic motion of scalar particles coupled to the Chern-Simons invariant in the Kerr spacetime \cite{Zhou} and stationary axisymmetric Einstein-Maxwell dilaton black hole spacetime \cite{Zhang}. Other generalized investigations based on the thermal chaos in the extended
phase space can be found, for example, in Refs. \cite{ChaosExtendedPhaseSpace1,ChaosExtendedPhaseSpace2,ChaosExtendedPhaseSpace3,ChaosExtendedPhaseSpace4}. 

Inspired by the aforementioned works, we will focus on the chaotic motion of particles around a Kerr black hole immersed in swirling universes in this work. Generally, the swirling universe refers to a rotating spacetime that is not asymptotically flat \cite{BarrientosCKMOP,BarrientosCHP}. From the Ernst formalism \cite{Ernst1,Ernst2}, Astorino \emph{et al.} constructed a new solution in Einstein's general relativity representing a Schwarzschild black hole immersed in swirling universes \cite{Astorino2}. This black hole spacetime is an algebraically general, stationary, axially symmetric and non-asymptotically flat black hole solution of the vacuum Einstein equations, with the special property that north and south hemispheres spin in opposite directions. As a further step, by using the Kerr metric in Boyer-Lindquist coordinates as a seed, they obtained a Kerr black hole immersed in swirling universes, which simultaneously possesses the swirling parameter and the spin parameter \cite{Astorino2}. In Ref. \cite{Capobianco}, Capobianco \emph{et al.} studied the geodesics of particles in a spacetime describing a swirling universe, and found that the geodesic equations can no longer be decoupled and rather small values of the swirling parameter will produce substantial changes with respect to the Schwarzschild orbits in general relativity. Moreira \emph{et al.} investigated the null geodesic flow and the existence of light rings of the black holes in swirling universes, and pointed out that the swirling parameter drives the light rings outside the equatorial plane and the contour of the shadow becomes a tilted oblate shape \cite{MoreiraHC}. More recently, the authors of Refs. \cite{ChenCJ} and \cite{Gjorgjieski} analyzed the geometrically thick equilibrium tori orbiting a Schwarzschild black hole in swirling universes and a Kerr black hole in swirling universes respectively, and obtained new effects from background swirling on the equilibrium tori. Thus, it would be of great interest to investigate the motion of particles in the spacetime of a Kerr black hole immersed in swirling universes. On the one hand, it is worthwhile to examine the influence of the swirling parameter on the motion of particles and identify whether the motion of particles are chaotic or not, due to the non-separability of the geodesic equations in a swirling universe \cite{Capobianco}. On the other hand, it would be important to see some general features for the motion of particles in a swirling universe and to check whether it is possible to distinguish between black holes with and without swirling parameters based on the (chaotic) behavior of dynamical systems. 

This work is organized as follows. In Section II, we briefly review the Kerr black hole immersed in swirling universes and give the geodesic equations of particles. In Section III, we investigate the chaotic motion of particles by using techniques including the Poincar\'{e} section, fast Lyapunov indicator, bifurcation diagram and basins of attraction, and probe the effects of swirling parameter and spin parameter on the chaotic behavior of particles in the spacetime of a Kerr black hole immersed in swirling universes. We will conclude in the last section with our main results.

\section{Geodesics of the particle in the spacetime of a Kerr black hole immersed in swirling universes}
	
The line element for the Kerr black hole solution in swirling universes, which is a stationary and axisymmetric spacetime by immersing the Kerr black hole into a rotating universe, can be expressed in Boyer-Lindquist coordinates as \cite{Astorino2}
\begin{equation}\label{metic}
	{ds}^2 = F (d\varphi - \omega dt)^2 + F^{-1}\biggl[-\rho^2 {dt}^2 + \Sigma\sin^2\theta \biggl(\frac{{dr}^2}{\Delta} + {d\theta}^2\biggr)\biggr],
\end{equation}
where the functions $F$ and $\omega$ are given by a finite power series of the swirling parameter $j$
\begin{equation}
	F^{-1} = X_{(0)} + jX_{(1)} + j^2 X_{(2)} , \qquad
	\omega = \omega_{(0)} + j\omega_{(1)} + j^2\omega_{(2)} ,
\end{equation}
with the expansion coefficients
\begin{subequations}
	\begin{align}
		X_{(0)} & = \frac{R^2}{\Sigma \sin^2\theta} \,, \\
		X_{(1)} & = \frac{4 a M \, \Xi \cos\theta }{\Sigma \sin^2\theta} \,, \\
		X_{(2)} & = \frac{4 a^2 M^2 \Xi^2 \cos^2\theta + \Sigma^2 \sin^4\theta}{R^2 \Sigma \sin^2\theta} \,,
	\end{align}
\end{subequations}
and 
\begin{subequations}
	\begin{align}
		\omega_{(0)} & = -\frac{2 a M r}{\Sigma}, \\
		\omega_{(1)} & = -\frac{4 \cos\theta [-a \Omega (r-M) + Ma^4 - r^4(r-2M) - \Delta a^2 r]}{\Sigma } \,, \\
		\omega_{(2)} & =
		-\frac{2M \{3ar^5 - a^5(r+2M) + 2a^3r^2(r+3M) - r^3(\cos^2\theta-6)\Omega + a^2[\cos^2\theta(3r-2M) - 6(r-M)] \Omega\}}{\Sigma } \,.
	\end{align}
\end{subequations}
Here we have defined 
\begin{subequations}
	\begin{align}
		\Delta & = r^2 - 2Mr + a^2 \,, \qquad\qquad\qquad\qquad\qquad\!\!\!\!\!\!\!\!
		\rho^2 = \Delta \sin^2\theta \,, \\
		\Sigma & = (r^2 + a^2)^2 - \Delta a^2 \sin^2\theta \,, \qquad\qquad\qquad\!\!\!\!\!\!
		\Omega = \Delta a \cos^2\theta \,, \\
		\Xi & = r^2(\cos^2\theta - 3) - a^2 (1+\cos^2\theta) \,, \qquad
		R^2 = r^2 + a^2 \cos^2\theta\,.
	\end{align}
\end{subequations}
with the mass parameter of the black hole $M$ and Kerr angular momentum per unit mass $a$. This metric reduces to the pure Kerr black hole when $j = 0$ and to the Schwarzschild black hole in swirling universes when $a=0$. As described in \cite{Astorino2}, the spin-spin interaction between the Kerr black hole and the background dragging creates a conical singularity along the symmetry axes and deforms the horizon geometry, and enhances the symmetry breaking regarding the spacetime properties, as shown for the horizons and ergosurfaces \cite{Gjorgjieski}. This transformed Kerr solution is non-asymptotically flat, and its north and south hemispheres spin in opposite directions. It should be noted that even small variations of $j$ alter the typically oblate structure of the Kerr horizon. 

From the Lagrangian of a timelike particle moving along the geodesic
\begin{equation}
\mathcal{L}=\frac{1}{2}g_{\mu\nu}\dot{x}^{\mu}\dot{x}^{\nu},
\end{equation}
for the metric (\ref{metic}) we obtain the geodesic equations of a particle
\begin{equation}\label{wfE1}
	\dot{t}=\frac{{g}_{\varphi\varphi} E+{g}_{t \varphi} L}{{g}_{t \varphi}^{2}-{g}_{t t} {g}_{\varphi\varphi}},\quad
	\dot{\varphi}=-\frac{{g}_{t \varphi} E+{g}_{t t} L}{{g}_{t \varphi}^{2}-{g}_{t t} {g}_{\varphi\varphi}},
\end{equation}
and
\begin{equation}\label{wfE2}
	\ddot{r}=\frac{1}{2} g^{rr}\left(g_{t t, r} \dot{t}^{2}-g_{r r, r} \dot{r}^{2}+g_{\theta \theta, r} \dot{\theta}^{2}+g_{\varphi\varphi, r} \dot{\varphi}^{2}+2 g_{t \varphi, r} \dot{t} \dot{\varphi}-2 g_{rr, \theta} \dot{r} \dot{\theta}\right),
\end{equation}
\begin{equation}\label{wfE3}
	\ddot{\theta}=\frac{1}{2} g^{\theta \theta}\left(g_{t t, \theta} \dot{t}^{2}+g_{r r, \theta} \dot{r}^{2}-g_{\theta \theta, \theta} \dot{\theta}^{2}+g_{\varphi\varphi, \theta} \dot{\varphi}^{2}+2 g_{t \varphi,\theta} \dot{t} \dot{\varphi}-2 g_{\theta \theta, r} \dot{r} \dot{\theta}\right),
\end{equation}
where $E$ and $L$ correspond to the energy and the angular momentum of a particle, respectively. Moreover, the motion of a particle also satisfies a constraint condition
\begin{equation}\label{Hcon}
	h=g_{t t} \dot{t}^{2}+g_{r r} \dot{r}^{2}+g_{\theta \theta} \dot{\theta}^{2}+g_{\varphi\varphi} \dot{\varphi}^{2}+2 g_{t \varphi} \dot{t} \dot{\varphi}+1=0.
\end{equation}
Obviously, the swirling parameter $j$ makes the differential Eq. (\ref{Hcon}) nonseparable, possibly resulting in the chaotic motion of particles, just as suggested in Ref. \cite{Capobianco}. Thus, in the next section we will scan the parameter space of the system for given values of $M$, $a$, $j$, $E$ and $L$, and search for the chaotic motion of particles in the spacetime of a Kerr black hole immersed in swirling universes.
	
\section{Chaotic motion of particles in the spacetime of a Kerr black hole immersed in swirling universes}
	
It is well known that the motion of particles in chaotic regions is highly sensitive to initial values, and the numerical method with high-precision is essential for solving the differential equations (\ref{wfE1})-(\ref{wfE3}) to avoid the pseudo-chaos produced by large numerical errors. Therefore, in this work we will use the corrected fifth-order Runge-Kutta method suggested in Refs. \cite{DZMa1, DZMa2},  where the velocities ($\dot{r}$, $\dot{\theta}$) are corrected in integration and the numerical deviation is pulled back in a least-squares shortest path.
	
Note that the motion of the particle is fully determined by its initial conditions and parameters in the system, but the choice for the parameters and initial conditions of the particle should be arbitrary in principle. Meanwhile, we find that it is nearly impossible to obtain the stable orbit of the particle with the high numerical value of $j$ in the numerical calculation in the natural unit ($G = c =\hbar=1$). Thus, for convenience, we choose the periodic orbit of the particle without swirling parameters as the initial motion orbit and investigate how the degree of disorder in the orbits changes with respect to the swirling parameter $j$ and the spin parameter $a$. Here we set the parameters $\{$$M=1$, $E=0.95$, $L=2.4M$$\}$ and initial conditions $\{$$ r(0)=7.2$, $\dot{r}(0)=0$,  $\theta(0)=\frac{\pi}{2}$$\}$ to obtain the desired regular orbit for 
the Schwarzschild black hole ($a = 0$) and Kerr black hole ($a = 0.54$) immersed in swirling universes. 

\begin{figure}[htbp!]
\includegraphics[width=5.5cm ]{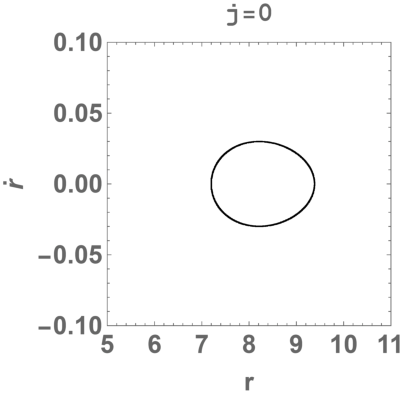}
\includegraphics[width=5.5cm ]{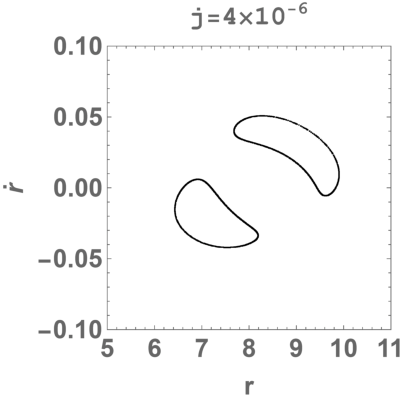}
\includegraphics[width=5.5cm ]{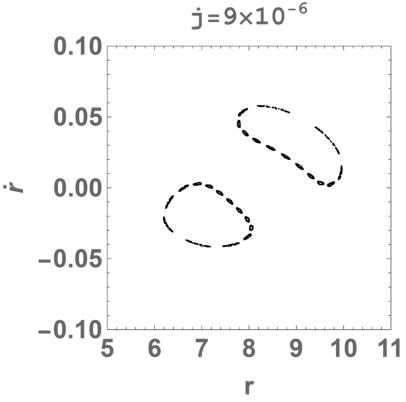}
\includegraphics[width=5.5cm ]{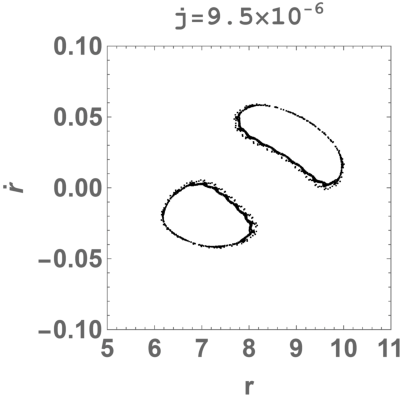}
\includegraphics[width=5.5cm ]{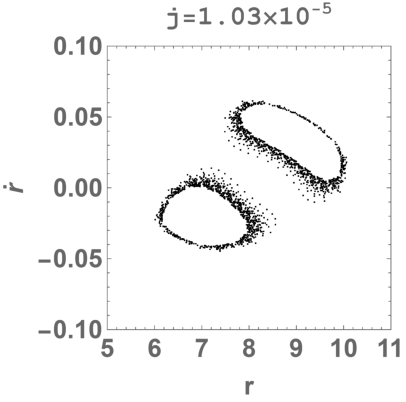}
\includegraphics[width=5.5cm ]{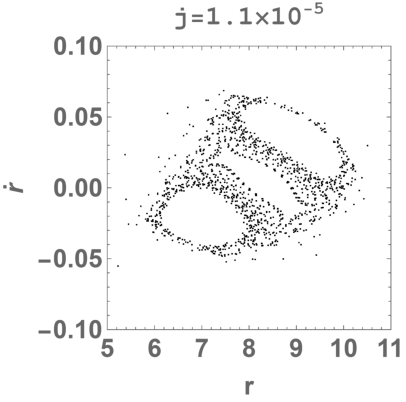}
\caption{The Poincar\'{e} section ($\theta = \frac{\pi}{2}$) with the swirling parameter $j$ for the motion of a particle in the Schwarzschild black hole ($a = 0$) immersed in swirling universes for the fixed values of $r(0) = 7.2$, $M = 1$, $E = 0.95$ and $L = 2.4M$. }\label{fig1}
\end{figure}

As a powerful tool for discerning the chaotic motions of particles, the Poincar\'{e} section is defined as the intersection of the trajectory within a continuous dynamical system with a given hypersurface which is transverse to the trajectory in the phase space. Based on the distribution of intersection points in the Poincar\'{e} section, the motions of the particle in the dynamical system can be classified into three types: the periodic motion, corresponding to a finite number of points; the quasi-periodic motion, corresponding to a series of close curves; the chaotic motion, corresponding to strange patterns of dispersed points with complex boundaries. For the Schwarzschild black hole immersed in swirling universes, i.e., $a=0$, we present the Poincar\'{e} sections ($r,\dot r$) with different swirling parameters $j$ for the motion of the particle in Fig. \ref{fig1}. When $j < 9.5\times10^{-6} $, the phase trajectories are quasi-periodic Kolmogorov–Arnold–Moser (KAM) tori and the behavior of the dynamical system is non-chaotic. Specifically, as the swirling parameter $j = 4 \times 10^{-6}$, there is an island chain consisting of two secondary KAM toris, which belong to the same trajectory. However, when $j \geq 9.5 \times 10^{-6}$, the KAM tori is destroyed and the corresponding trajectory becomes non-integrable, indicating the chaotic behavior of the system. As $j\rightarrow 1.1 \times 10^{-5}$, there exist a few discrete points in the Poincar\'{e} section because the particle eventually falls into the event horizon or escapes to the spatial infinity. So the swirling parameter $j$ makes the motion of particles more complex.

\begin{figure}[htbp!]
\includegraphics[width=5.5cm ]{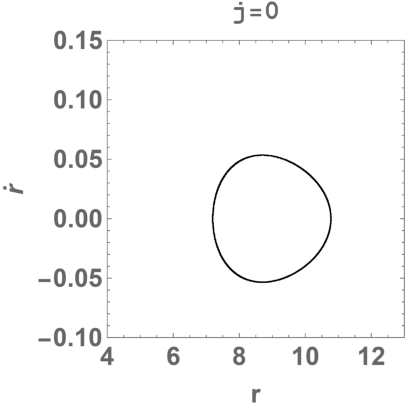}
\includegraphics[width=5.5cm ]{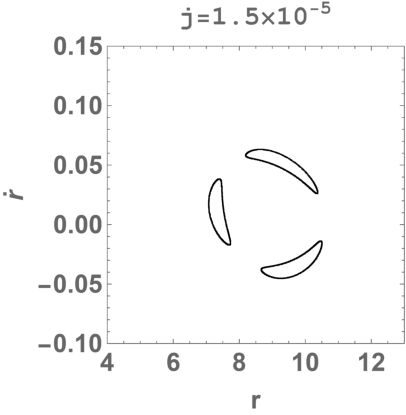}
\includegraphics[width=5.5cm ]{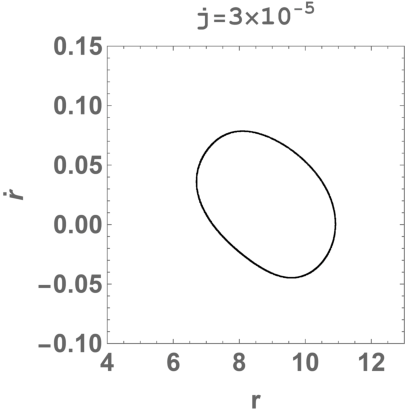}
\includegraphics[width=5.5cm ]{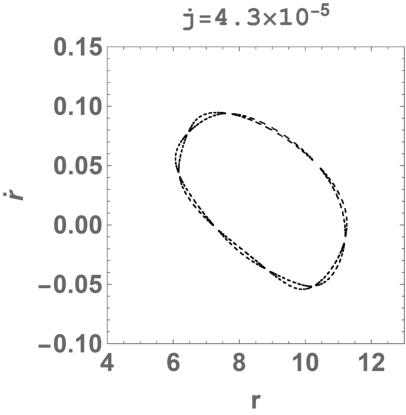}
\includegraphics[width=5.5cm ]{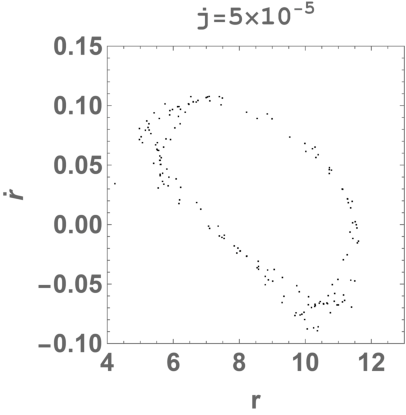}
\includegraphics[width=5.5cm ]{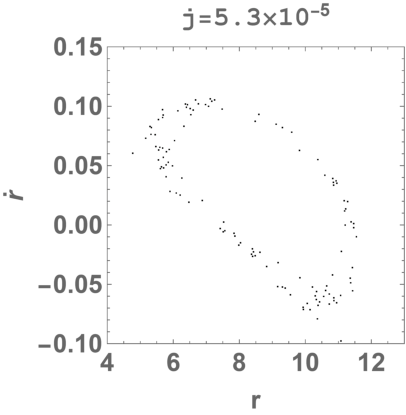}
\caption{The Poincar\'{e} section ($\theta = \frac{\pi}{2}$) with the swirling parameter $j$ for the motion of a particle in the Kerr black hole ($a=0.54$) immersed in swirling universes for the fixed values of $r(0) = 7.2$, $M = 1$, $E = 0.95$ and $L = 2.4M$. }\label{fig2}
\end{figure}

For the Kerr black hole immersed in swirling universes, i.e., $a=0.54$, Fig. \ref{fig2} shows the Poincar\'{e} sections with different swirling parameters $j$ for the motion of the particle. Similar to the Schwarzschild case in Fig. \ref{fig1}, we observe that the non-integrability of the motion of the particle increases as the swirling parameter $j$ increases for the Kerr black hole immersed in swirling universes. Interestingly, we find that there is an island chain consisting of three secondary KAM toris belonging to the same trajectory when $j = 1.5 \times 10^{-5}$, which is different from that of the Schwarzschild black hole immersed in swirling universes when $j = 4 \times 10^{-6}$, where there is an island chain consisting of two secondary KAM toris. Moreover, as the swirling parameter $j\rightarrow 3 \times 10^{-5}$, the chains of islands are joined together and become a big KAM tori, which indicates that the trajectory of this case is regular and integrable. Thus, the combination of the swirling parameter $j$ and the spin parameter $a$ provides richer physics in the motion of particles.

\begin{figure}[htbp!]
\includegraphics[width=5.5cm ]{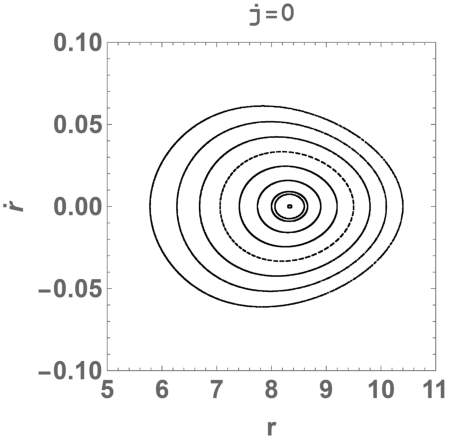}
\includegraphics[width=5.5cm ]{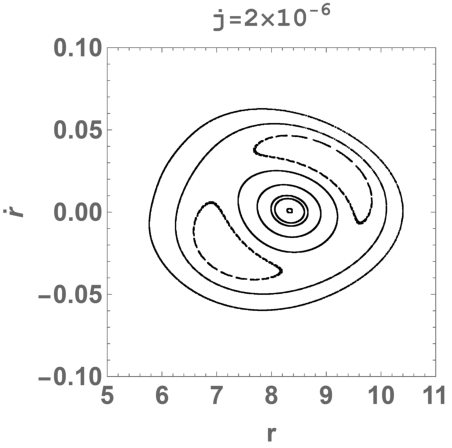}
\includegraphics[width=5.5cm ]{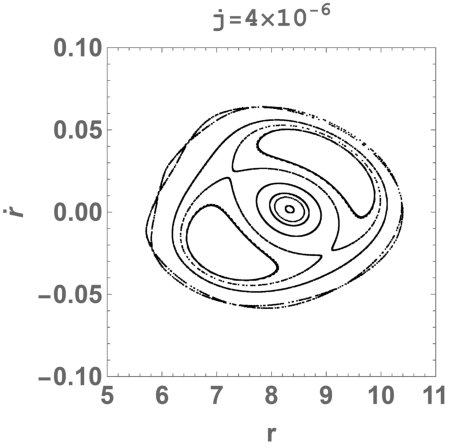}
\includegraphics[width=5.5cm ]{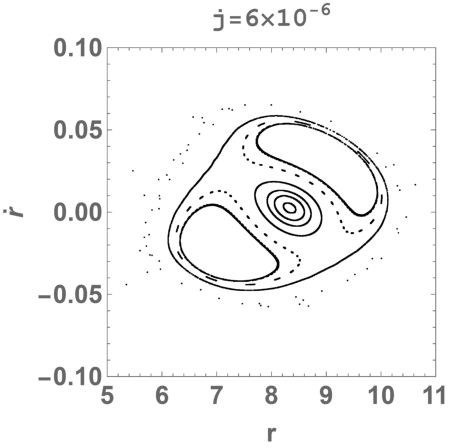}
\includegraphics[width=5.5cm ]{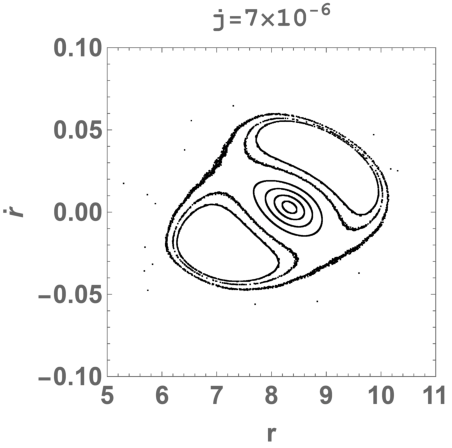}
\includegraphics[width=5.5cm ]{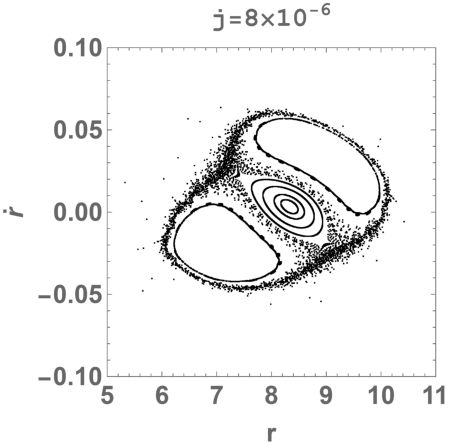}
\includegraphics[width=5.5cm ]{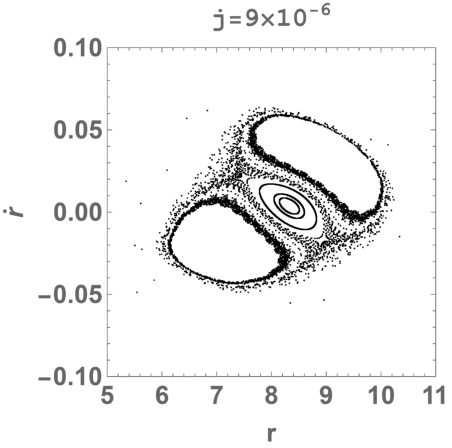}
\includegraphics[width=5.5cm ]{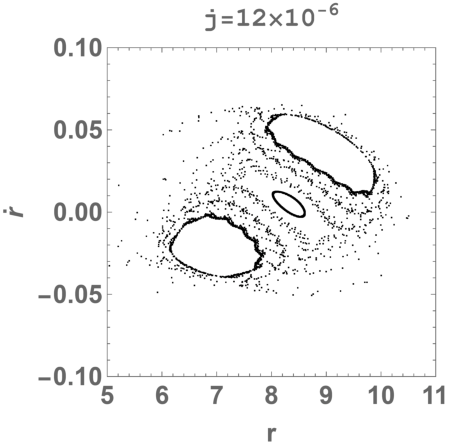}
\includegraphics[width=5.5cm ]{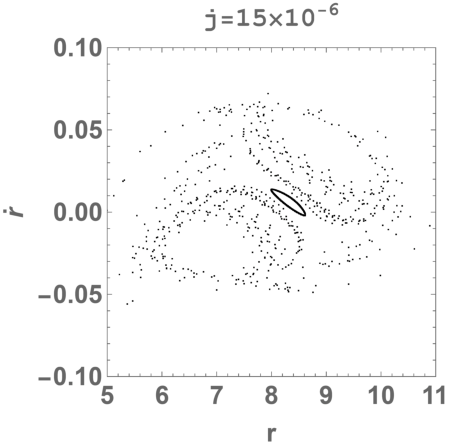}
\caption{The Poincar\'{e} section ($\theta = \frac{\pi}{2}$) with the swirling parameter $j$ for the motion of a particle in the Schwarzschild black hole ($a = 0$) immersed in swirling universes for the fixed values of $M = 1$, $E = 0.95$ and $L = 2.4M$.}\label{fig3}
\end{figure}

\begin{figure}[htbp!]
\includegraphics[width=5.5cm ]{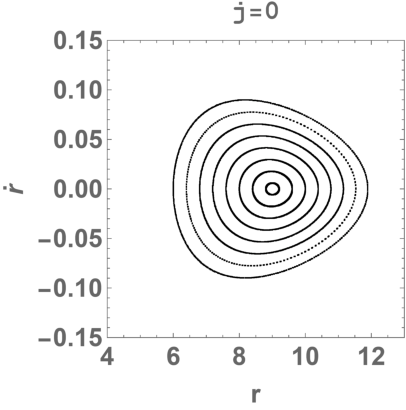}
\includegraphics[width=5.5cm ]{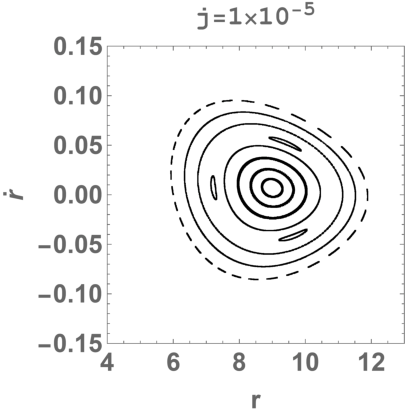}
\includegraphics[width=5.5cm ]{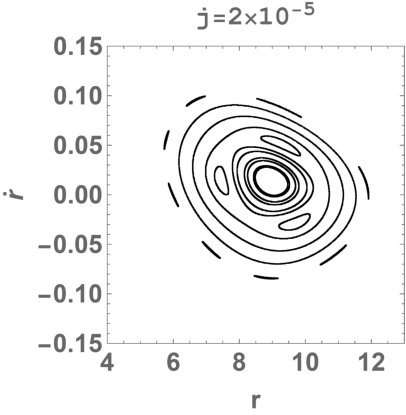}
\includegraphics[width=5.5cm ]{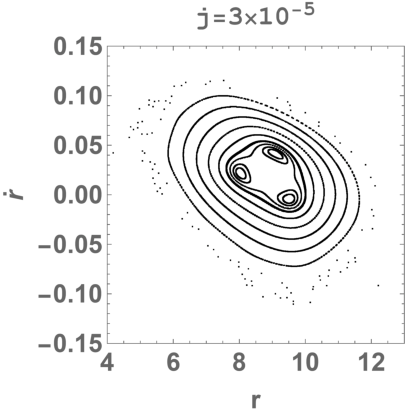}
\includegraphics[width=5.5cm ]{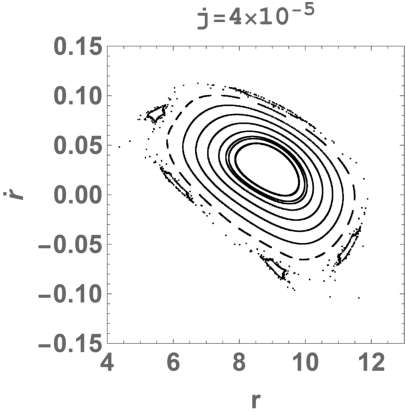}
\includegraphics[width=5.5cm ]{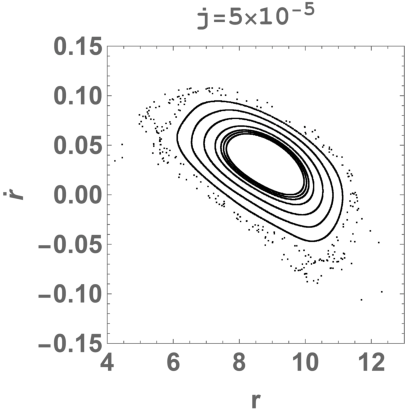}
\includegraphics[width=5.5cm ]{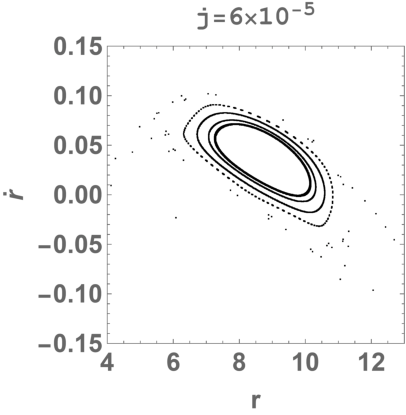}
\includegraphics[width=5.5cm ]{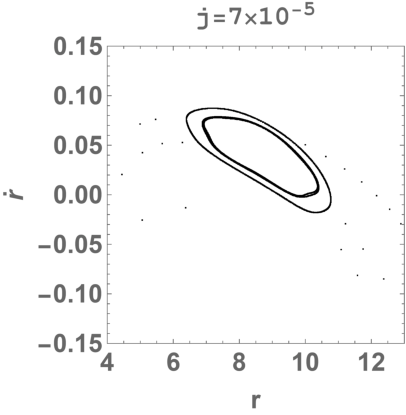} 
\includegraphics[width=5.5cm ]{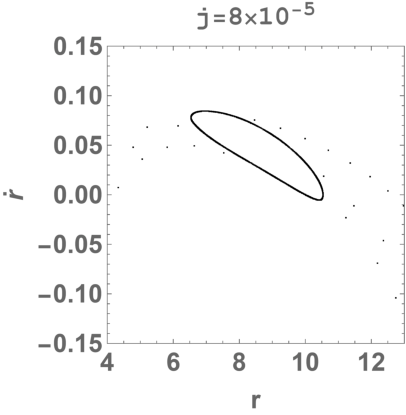}     
\caption{The Poincar\'{e} section ($\theta = \frac{\pi}{2}$) with the swirling parameter $j$ for the motion of a particle in the Kerr black hole ($a = 0.54$) immersed in swirling universes for the fixed values of $M = 1$, $E = 0.95$ and $L = 2.4M$. }\label{fig4}
\end{figure}

\begin{figure}[htbp!]
\includegraphics[width=5.5cm ]{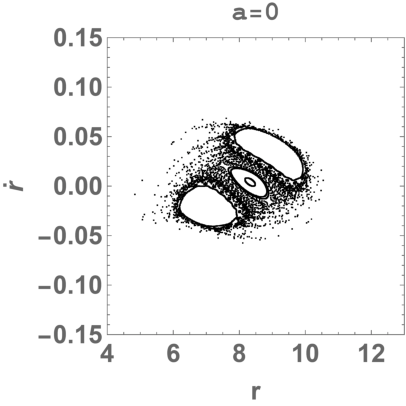}
\includegraphics[width=5.5cm ]{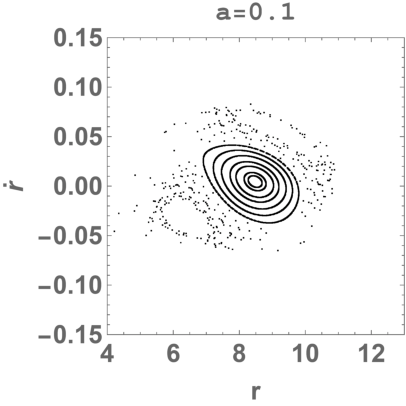}
\includegraphics[width=5.5cm ]{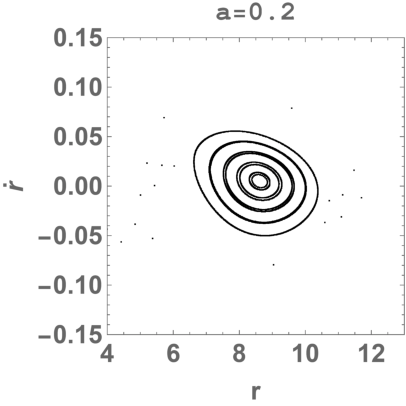}
\includegraphics[width=5.5cm ]{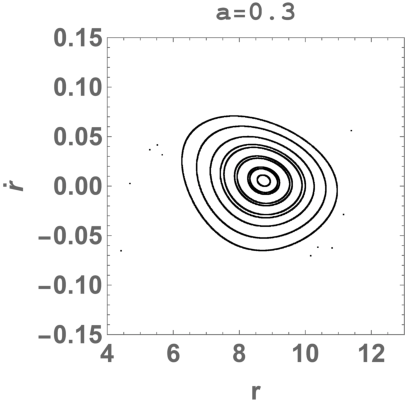}
\includegraphics[width=5.5cm ]{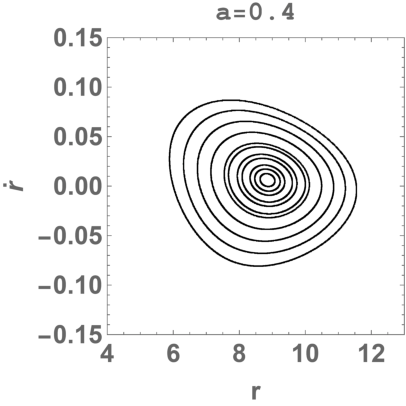}
\includegraphics[width=5.5cm ]{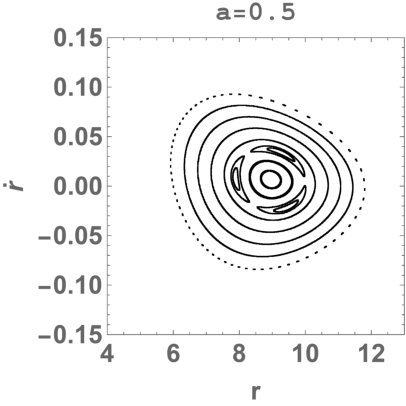}
\includegraphics[width=5.5cm ]{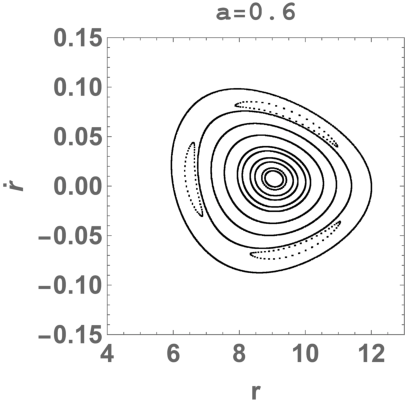}
\includegraphics[width=5.5cm ]{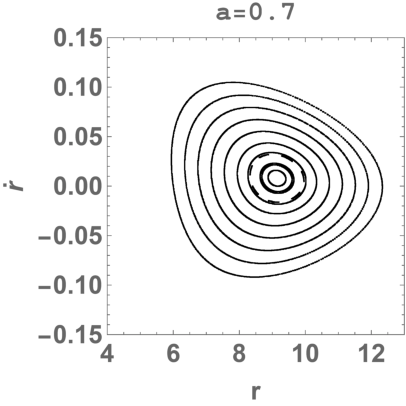}
\includegraphics[width=5.5cm ]{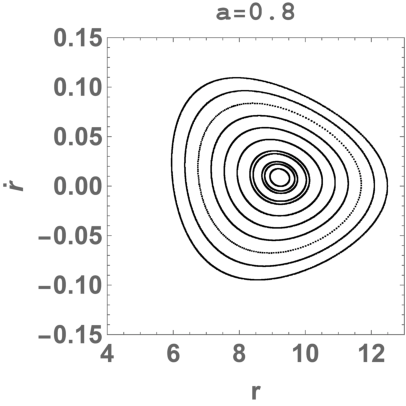}
\caption{The Poincar\'{e} section ($\theta = \frac{\pi}{2}$) with the spin parameter $a$ for the motion of a particle in the Kerr black hole immersed in swirling universes for the fixed values of $j =1 \times 10^{-5}$, $M = 1$, $E = 0.95$ and $L = 2.4M$.}\label{fig5}
\end{figure}

In Figs. \ref{fig3} - \ref{fig5}, we exhibit the Poincar\'{e} sections with more motion orbits of the particle in the spacetimes of Schwarzschild black hole and Kerr black hole immersed in swirling universes. It is found that the chaotic motion of the particle mainly occurs in regions where the swirling parameter $j$ takes relatively small values in the swirling blackground. In Figs. \ref{fig3} and \ref{fig4}, the Poincar\'{e} sections consist of a series of closed curves for $j=0$, meaning that all orbits are regular because the motion equations of the particle reduce to the usual variable-separable geodesic equation. As the swirling parameter $j$ increases, the main island of stability shrinks and the chaotic region increases. However, the chaotic intensity of chaotic orbits for $a=0$ is stronger than that for $a=0.54$, as verified by Fig. \ref{fig5}, which shows that both the chaotic region and intensity decrease with the increase of the spin parameter $a$ for the fixed swirling parameter $j = 10^{-5}$. Therefore, the spin parameter $a$ and the swirling parameter $j$ have completely different effects on the motion of particles, i.e., the non-integrability of the motion of particles decreases as $a$ increases but increases as $j$ increases.

\begin{figure}[htbp!]
\includegraphics[width=5.3cm ]{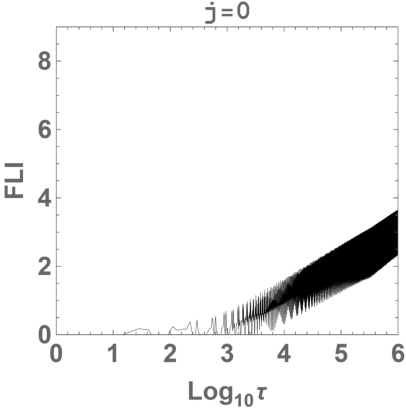}
\includegraphics[width=5.3cm ]{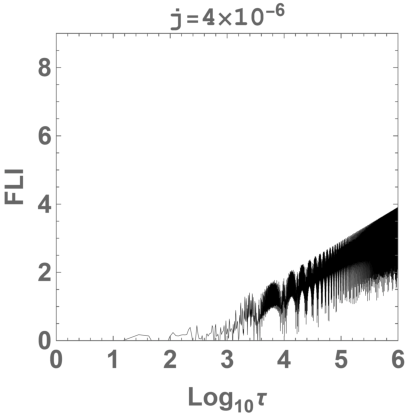}
\includegraphics[width=5.3cm ]{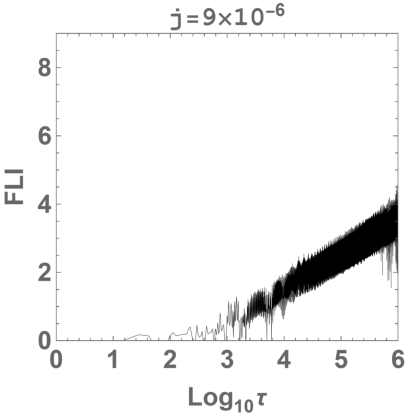}
\includegraphics[width=5.3cm ]{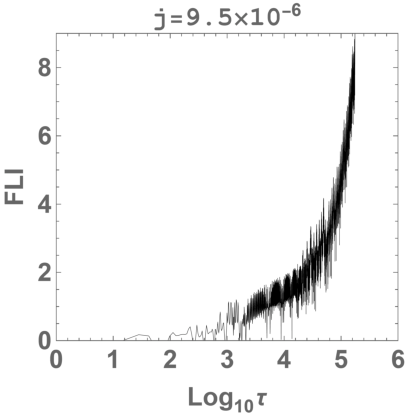}
\includegraphics[width=5.3cm ]{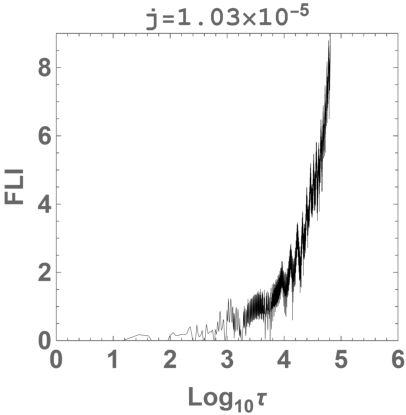}
\includegraphics[width=5.3cm ]{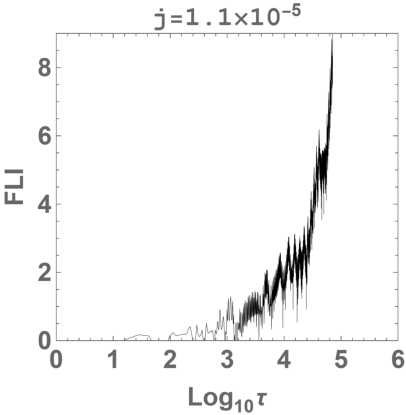}
\caption{The fast Lyapunov indicator (FLI) with the swirling parameter $j$ for the signals shown in Fig. \ref{fig1}. }\label{fig6}
\end{figure}

\begin{figure}[htbp!]
\includegraphics[width=5.3cm ]{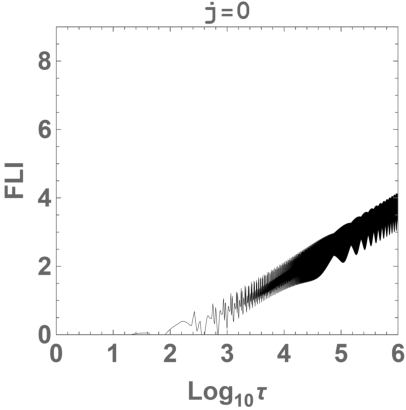}
\includegraphics[width=5.3cm ]{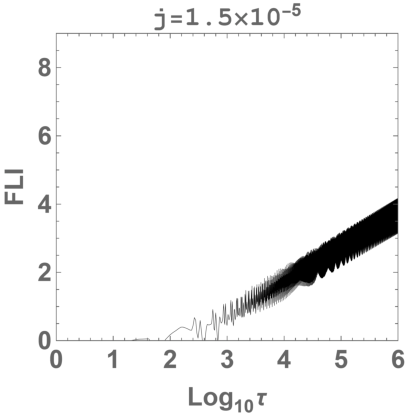}
\includegraphics[width=5.3cm ]{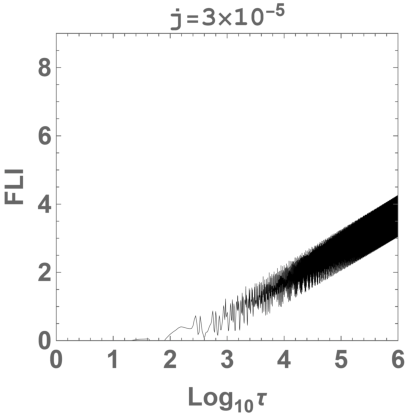}
\includegraphics[width=5.3cm ]{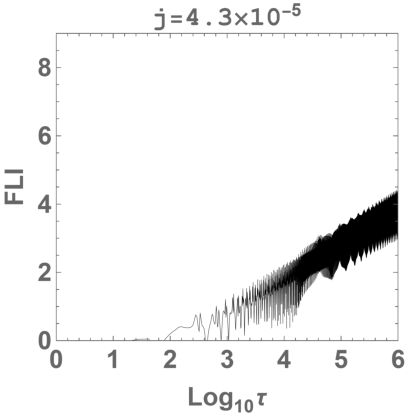}
\includegraphics[width=5.3cm ]{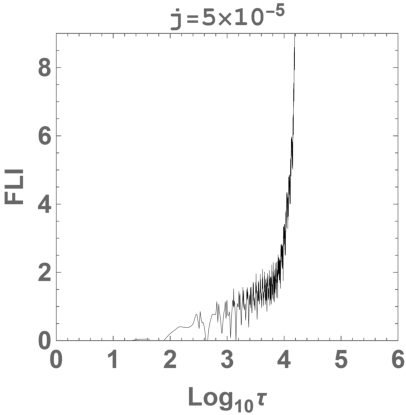}
\includegraphics[width=5.3cm ]{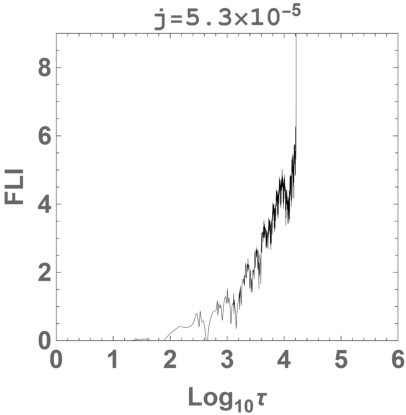}
\caption{The fast Lyapunov indicator (FLI) with the swirling parameter $j$ for the signals shown in Fig. \ref{fig2}. }\label{fig7}
\end{figure}

The fast Lyapunov indicator (FLI) is another efficient tool to identify the chaotic orbits of particles by measuring two adjacent orbits over time with the average separation index. In the curved spacetime, we can express the FLI with two-particle method as \cite{Tancredi,Froe,Wu,Chen}
\begin{eqnarray}
FLI(\tau)=-(k+1)\ast\log_{10}d(0)+\log_{10}d(\tau),
\end{eqnarray}
where $d(\tau)=\sqrt{|g_{\mu\nu}\Delta x^{\mu}\Delta x^{\nu}|}$ is the distance between the two particles with the deviation vector $\Delta x^{\mu}$ between two nearby trajectories, and $k$ is the sequential number of renormalization used to avoid numerical saturation arising from the rapid separation of these two trajectories. The $\text{FLI}(\tau)$ grows algebraically with time for the regular or periodic orbit, but grows exponentially for the chaotic orbit. In Figs. \ref{fig6} and \ref{fig7} , we give the $\text{FLI}(\tau)$ with the swirling parameter $j$ for the selected initial orbit presented in Figs. \ref{fig1} and \ref{fig2}. It is shown that, with the increase of time $\tau$, the $\text{FLI}(\tau)$ grows with exponential rate for the signals when $j \geq 9.5 \times 10^{-6}$ in Fig. \ref{fig6} and $j \geq 5 \times 10^{-5}$ in Fig. \ref{fig7}, and the corresponding motions are chaotic. But when $j < 9.5 \times 10^{-6}$ in Fig. \ref{fig6} and $j < 5 \times 10^{-5}$ in Fig. \ref{fig7}, the $\text{FLI}(\tau)$ increases linearly with $\tau$, and so the motions of the particle are regular. These findings are in good agreement with those obtained from the Poincar\'{e} sections shown in Figs. \ref{fig1} and \ref{fig2}.

\begin{figure}[htbp!]
\includegraphics[width=5.3cm ]{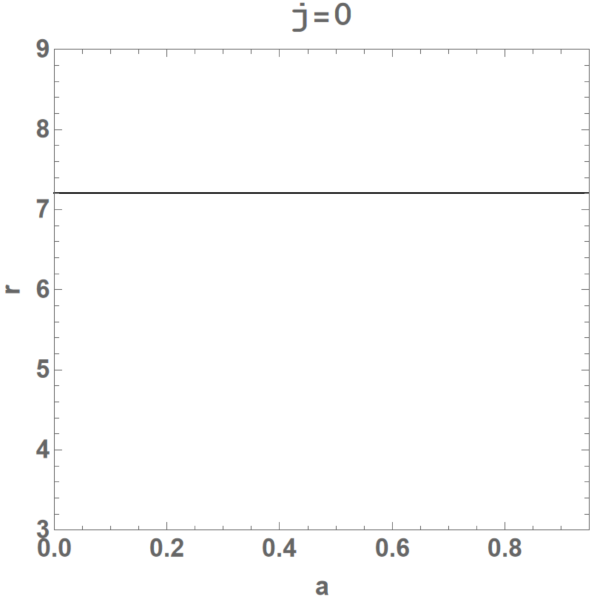}
\includegraphics[width=5.3cm ]{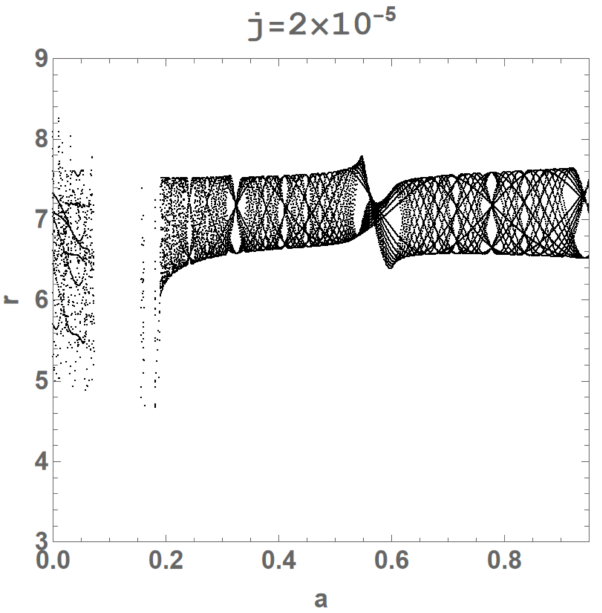}
\includegraphics[width=5.3cm ]{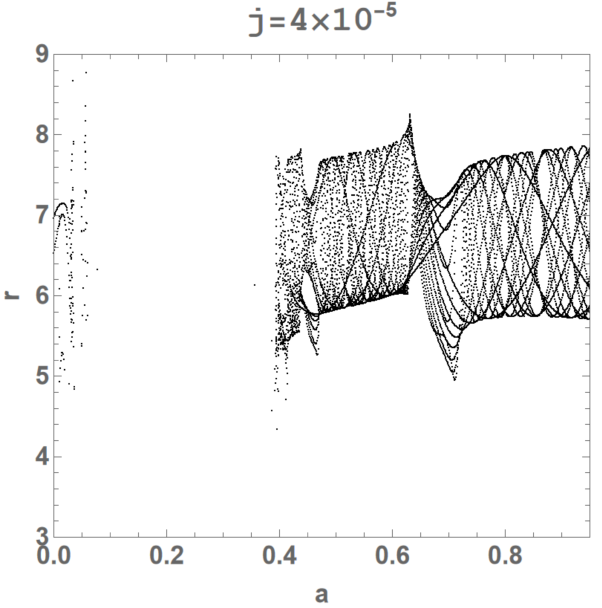}
\includegraphics[width=5.3cm ]{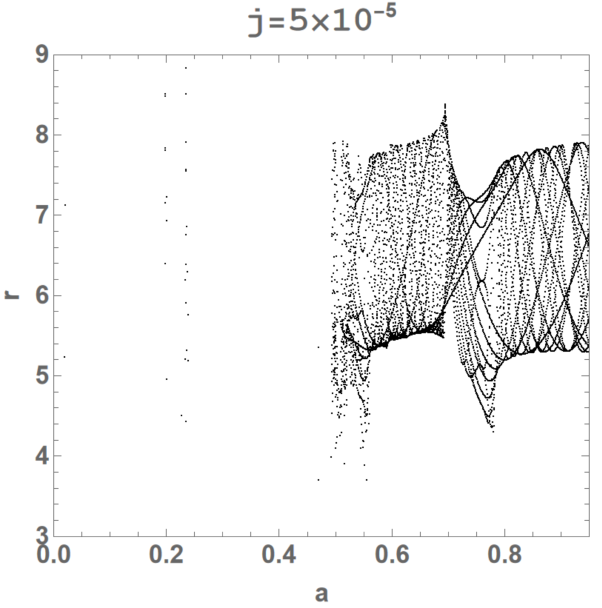}
\includegraphics[width=5.3cm ]{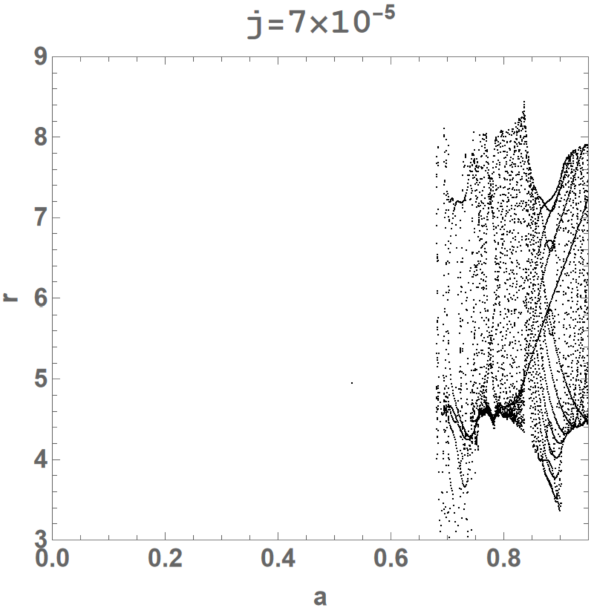}
\includegraphics[width=5.3cm ]{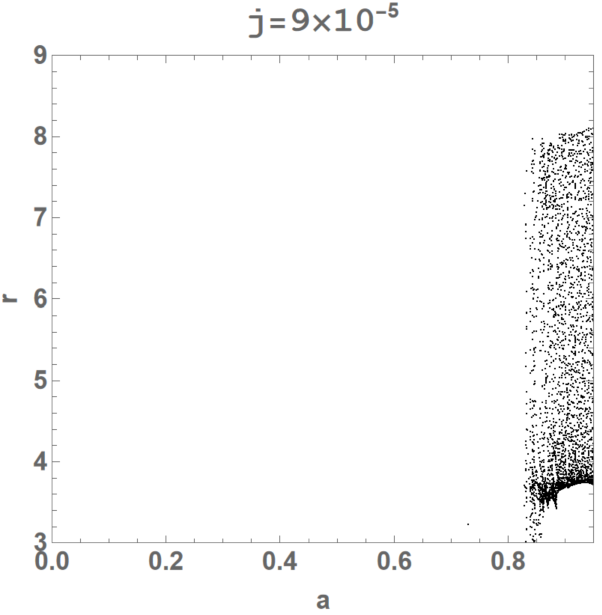}
\caption{The bifurcation changes with the spin parameter $a$ for different values of the swirling parameter $j$. }\label{fig8}
\end{figure}

\begin{figure}[htbp!]
\includegraphics[width=5.5cm ]{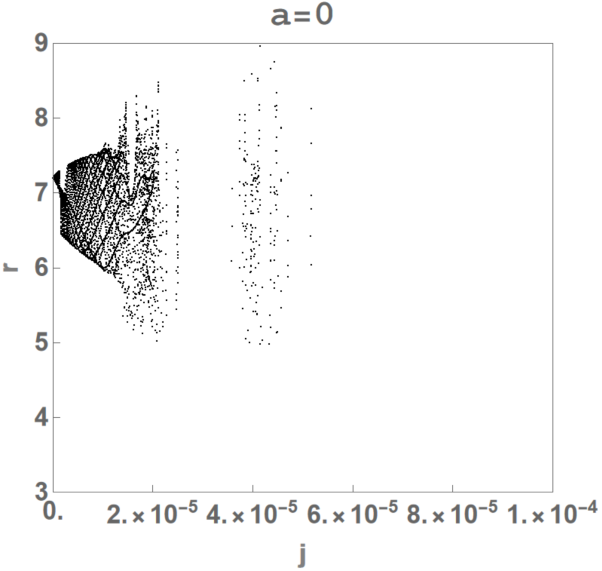}
\includegraphics[width=5.5cm ]{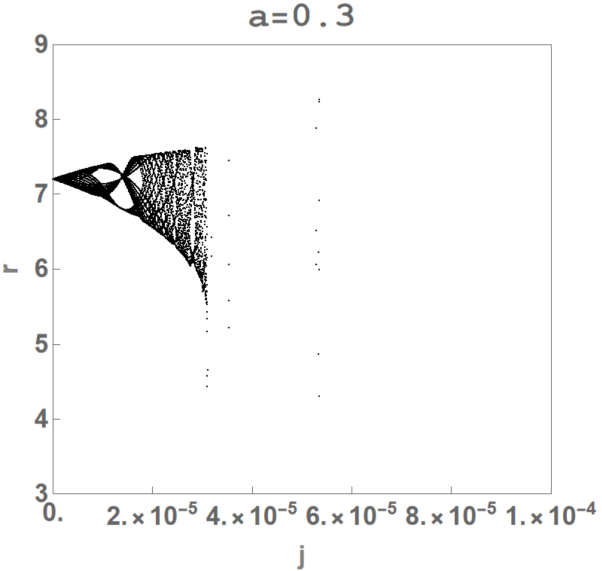}
\includegraphics[width=5.5cm ]{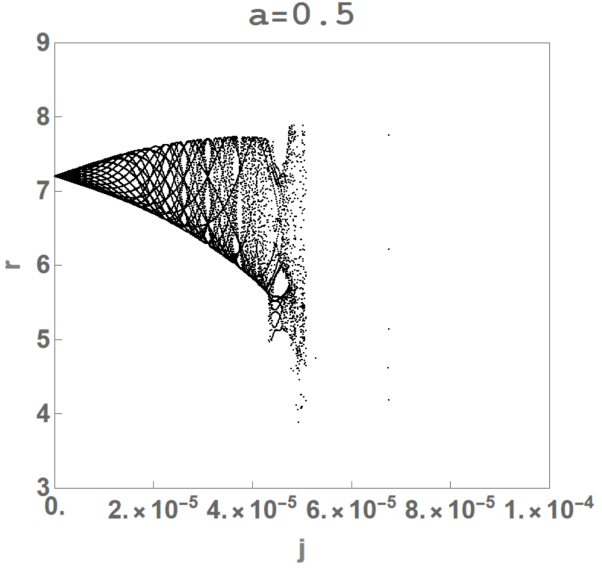}
\includegraphics[width=5.5cm ]{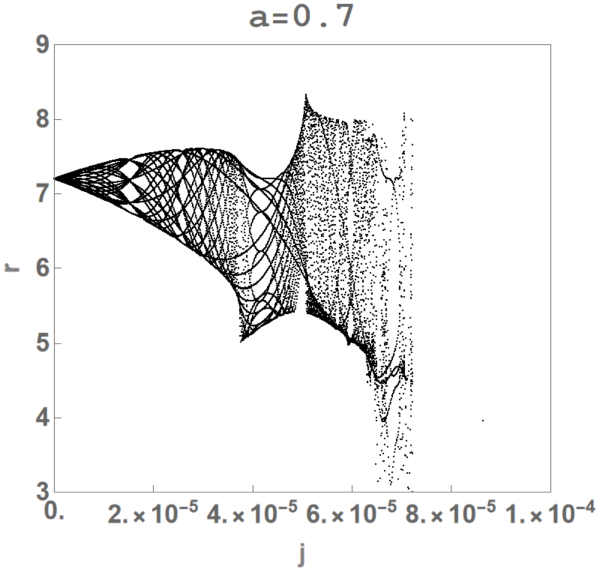}
\includegraphics[width=5.5cm ]{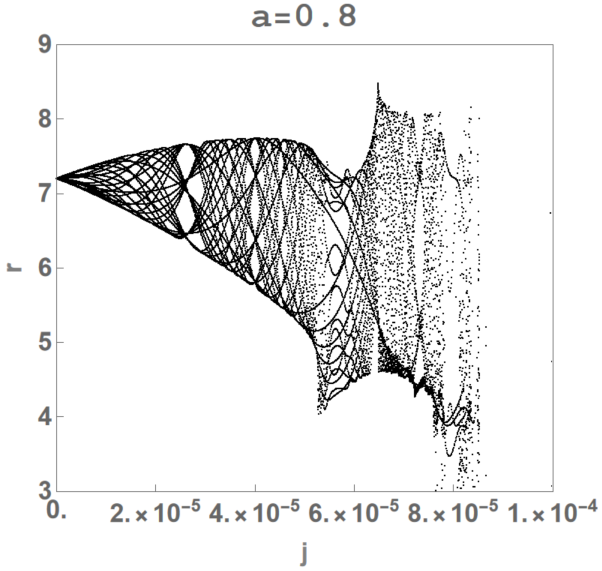}
\includegraphics[width=5.5cm ]{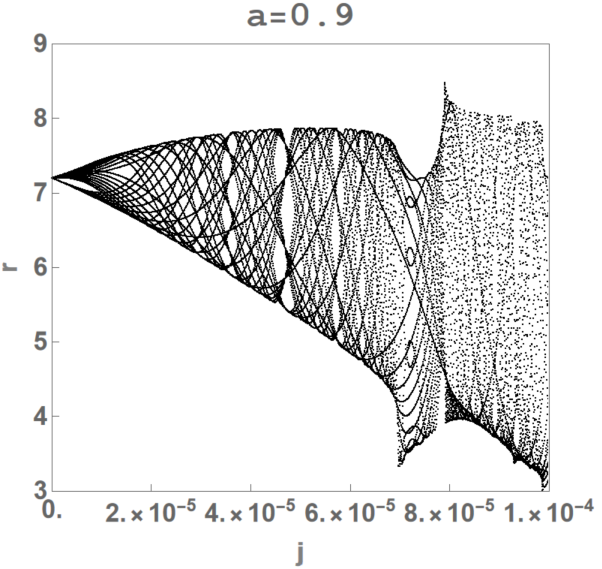}
\caption{The bifurcation changes with the swirling parameter $j$ for different values of the spin parameter $a$. }\label{fig9}
\end{figure}

Now we move to the bifurcation diagram, which can show us how the dynamical behaviors of system depend on the black hole parameters. In Figs. \ref{fig8} and \ref{fig9}, we plot the bifurcation diagrams of the radial coordinate of the particle with the swirling parameter $j$ and the spin parameter $a$ in the spacetime of a Kerr black hole immersed in swirling universes, and investigate the effects of $j$ and $a$ on the motion of particles. When the swirling parameter $j=0$, the radial coordinate $r(\tau)$ is a periodic function and there is no bifurcation for the dynamical system, just as shown in the first panel in Fig. \ref{fig8}, which suggests that the motions of particles are regular in this case. From Fig. \ref{fig8}, we see that as the swirling parameter $j$ increases, both the lower bound of $a$ for chaotic orbits and the range of $r$ in the chaotic solutions increase,  which shows that the presence of $j$ changes the range of $a$ where the chaotic motion appears for particles. Similarly, from Fig. \ref{fig9}, both the lower bound of $j$ for chaotic orbits and the range of $r$ in the chaotic solutions increase with the increase of $a$. These results indicate that the swirling parameter $j$ and the spin parameter $a$ yield much richer effects on the motion of particles.

\begin{figure}[htbp!]
\includegraphics[width=3.98cm ]{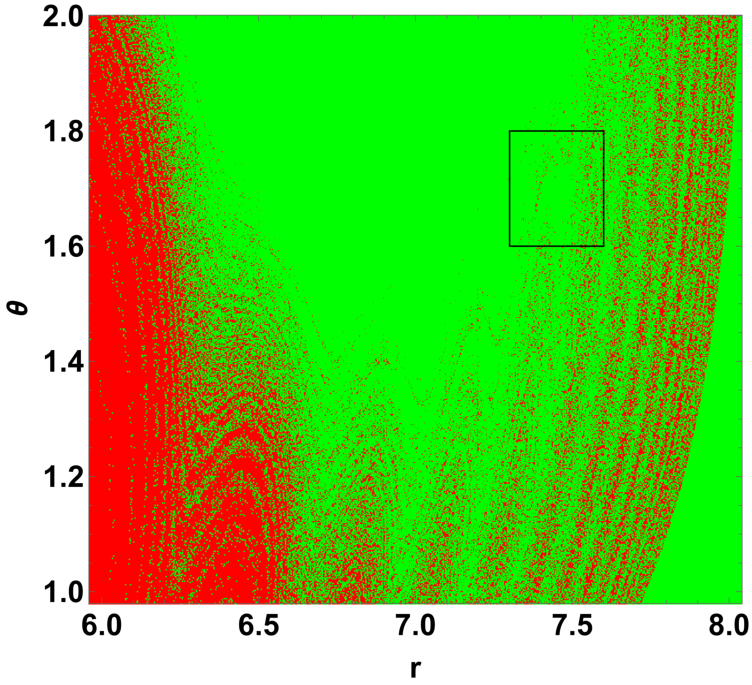}
\includegraphics[width=4.08cm ]{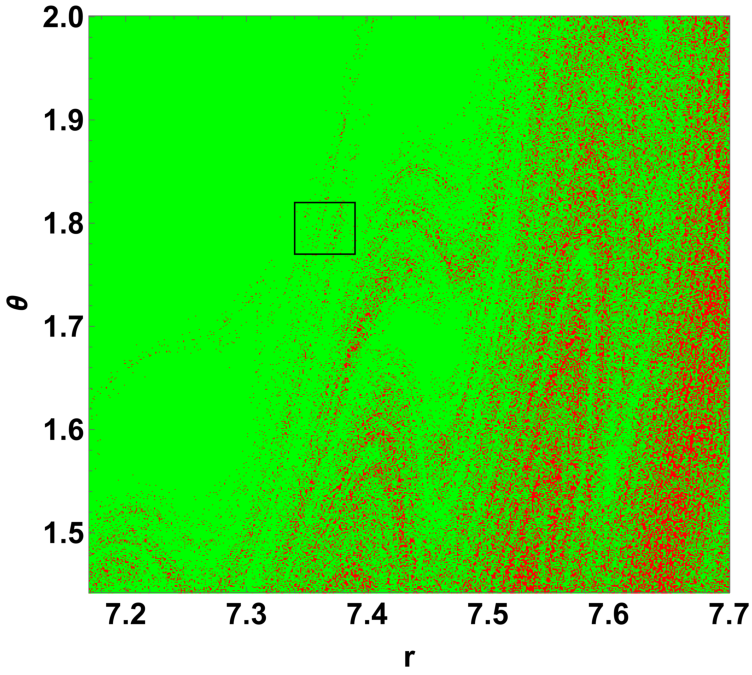}
\includegraphics[width=4.25cm ]{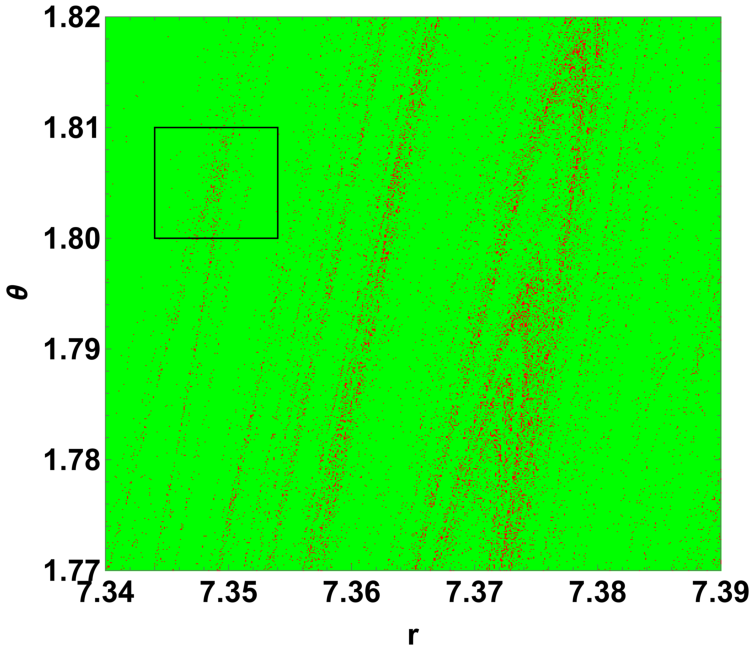}
\includegraphics[width=4.1cm ]{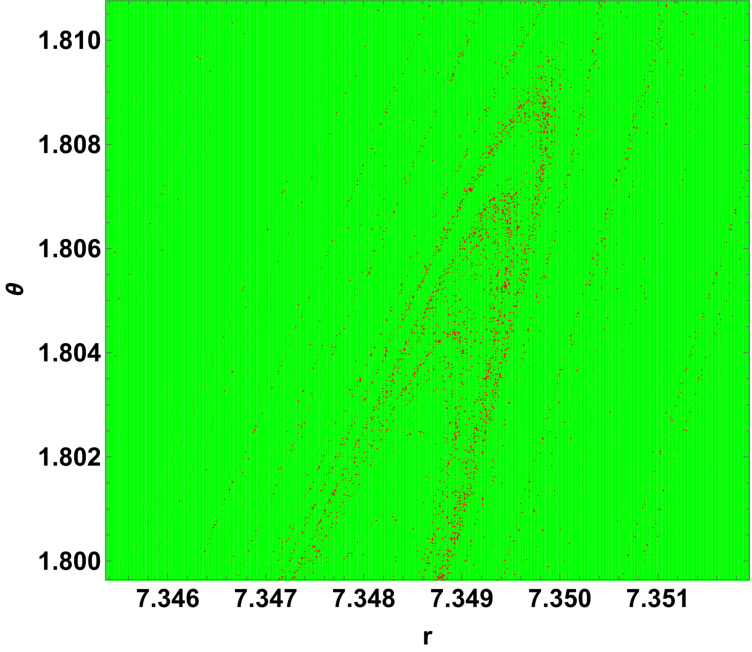}
\caption{The fractal basins of attraction for the particle in the Schwarzschild black hole ($a = 0$) immersed in swirling universes with the fixed parameters $j =1 \times 10^{-5}$, $M = 1$, $E = 0.95$ and $L = 2.4M$. }\label{fig10}
\end{figure}

\begin{figure}[htbp!]
\includegraphics[width=3.9cm]{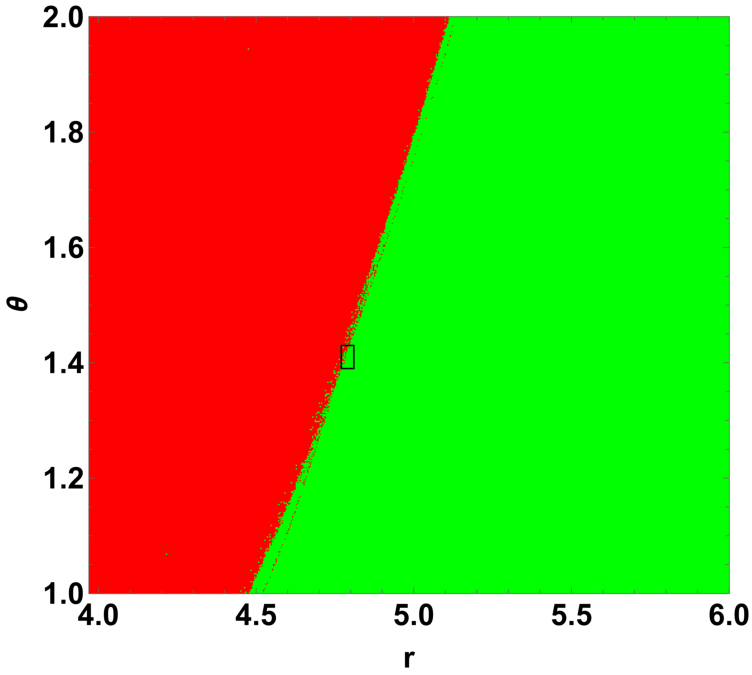}
\includegraphics[width=4.2cm]{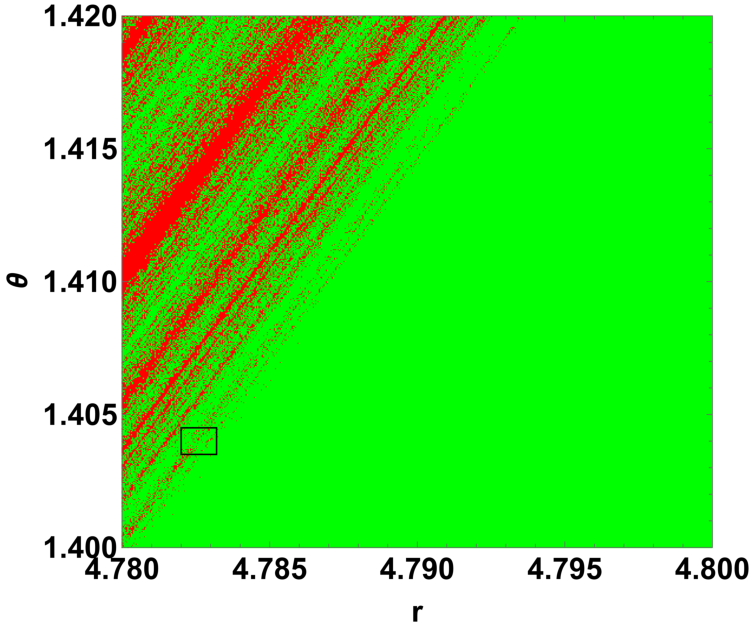}
\includegraphics[width=4.1cm]{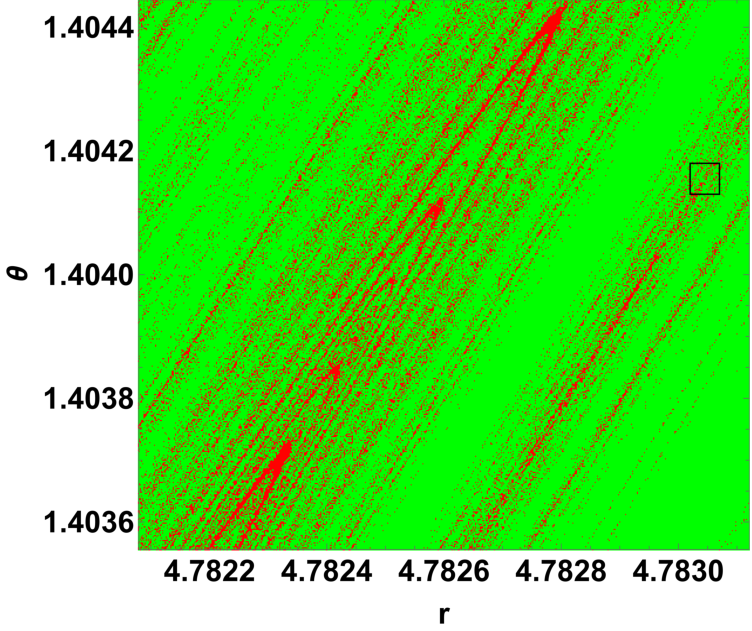}
\includegraphics[width=4.2cm]{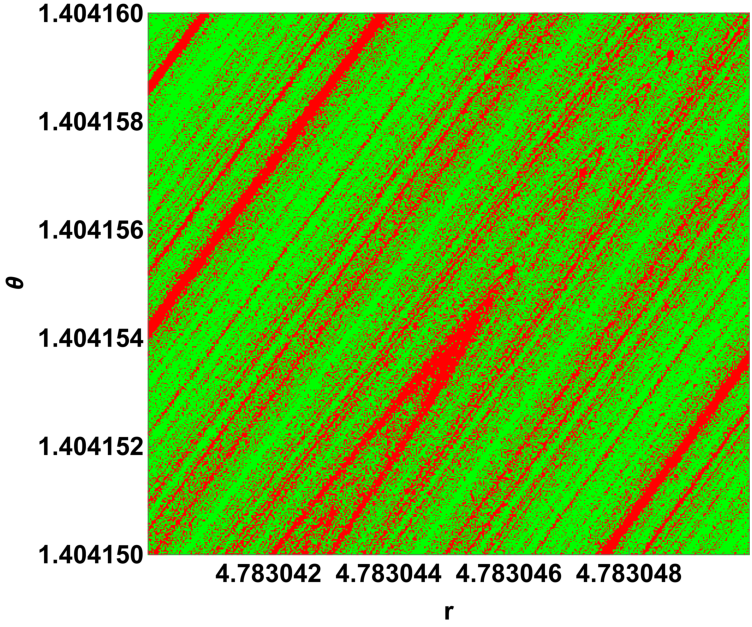}
\caption{The fractal basins of attraction for the particle in the Kerr black hole ($a = 0.54$) immersed in swirling universes with the fixed parameters $j =1 \times 10^{-5}$, $M = 1$, $E = 0.95$ and $L = 2.4M$. }\label{fig11}
\end{figure}

\begin{figure}[htbp!]
\includegraphics[width=3.8cm]{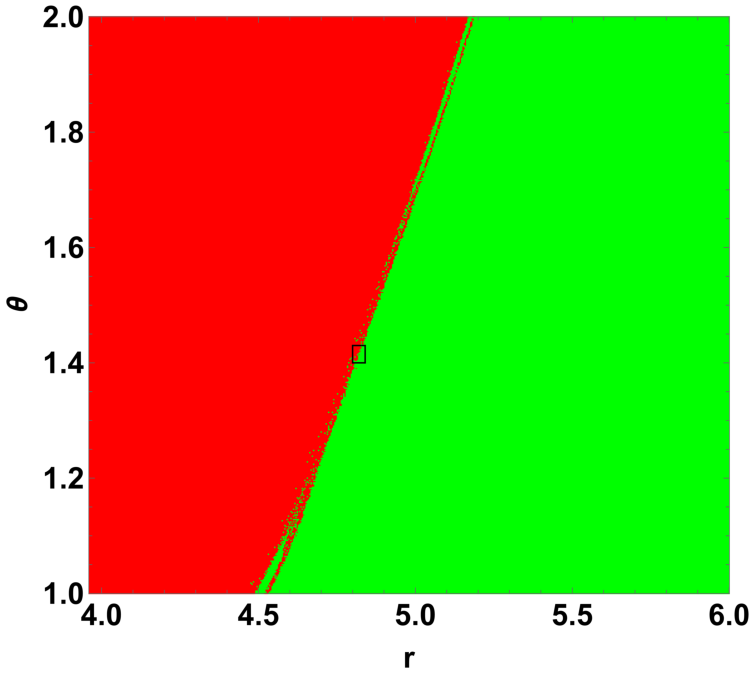}
\includegraphics[width=4.15cm]{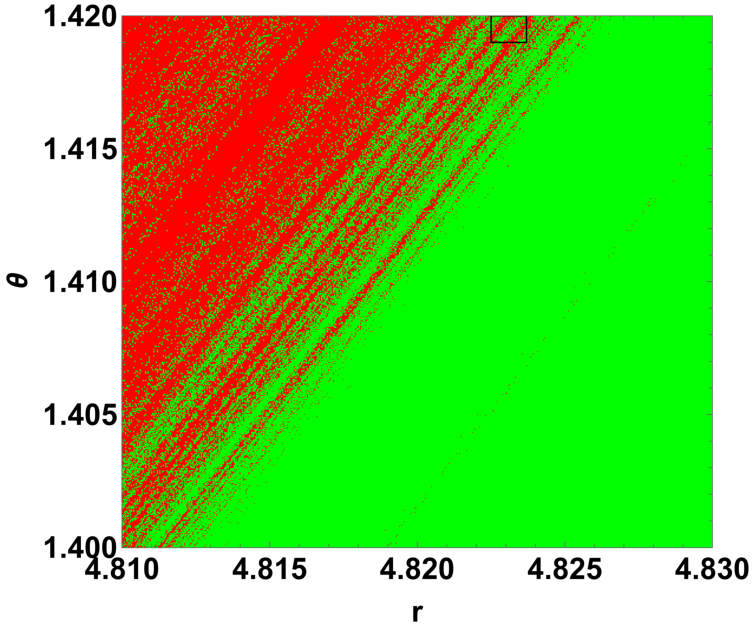}
\includegraphics[width=4.33cm]{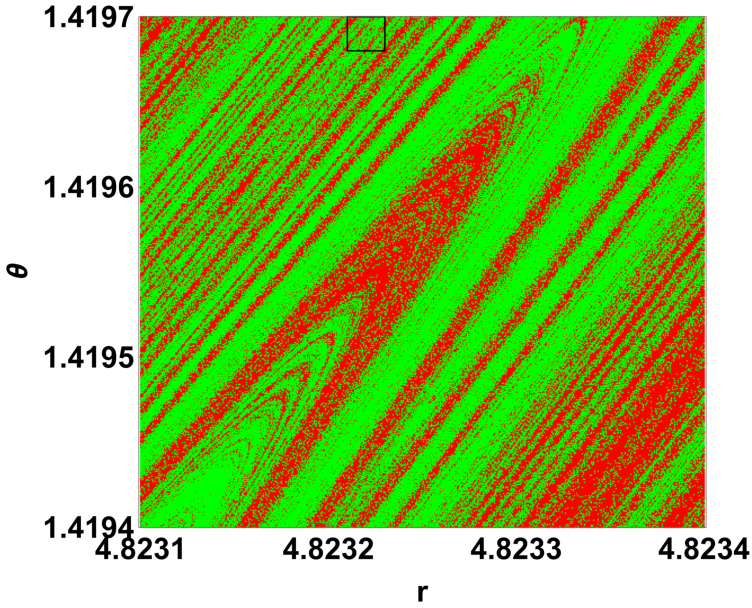}
\includegraphics[width=4.15cm]{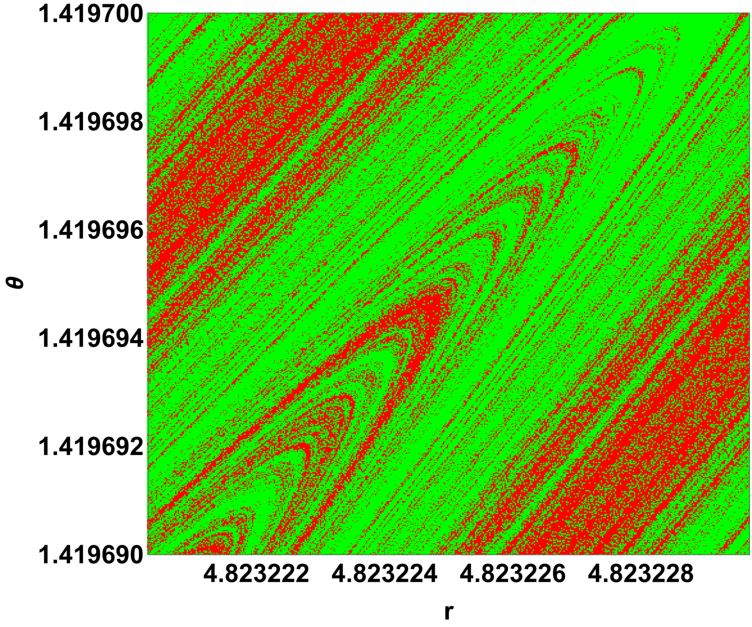}
\caption{The fractal basins of attraction for the particle in the Kerr black hole ($a = 0.54$) immersed in swirling universes with the fixed parameters $j =1.1 \times 10^{-5}$, $M = 1$, $E = 0.95$ and $L = 2.4M$. }\label{fig12}
\end{figure}

Finally, we consider the basin boundaries of attractors, which can provide a signature of chaos \cite{DettmannFC,FrolovL,ZayasT,Basin}. Actually, the smooth basin boundaries indicate the regular dynamics, while the fractal boundaries suggest the chaotic motions of orbits. In Figs. \ref{fig10} - \ref{fig12}, we show the basins of attraction in a large subset of phase space for particles in the spacetimes of Schwarzschild black hole and Kerr black hole immersed in swirling universes with the fixed parameters $M = 1$, $E = 0.95$ and $L = 2.4M$, where the initial conditions correspond to the points in these figures are set to $\dot{r} = 0$ with $\dot{\theta}$ given by the constraint (\ref{Hcon}), i.e., $h = 0$. The red points denote the case of geodesics crossing into the event horizon $r_{H}$, where the particle falls into the black hole along geodesics, the blue points correspond to the particle escaping to infinity and the green points represent the particle oscillating around the black hole. In our investigation, the condition for capture is $r \leq r_{H}$ and the condition for escape is $r \geq 100r_{H}$. For green points, we consider trajectories that neither got captured nor escaped to infinity after 100,000 iterations. From these figures, we can observe clearly that, regardless of the spin parameter $a$, there exist some self-similar fractal fine structures in the basins boundaries of attractors, which implies that there exists the chaotic motion for a particle in the spacetime of a black hole immersed in swirling universes.

\section{Summary}
	
We have investigated the motion of particles in the spacetime of a Kerr black hole immersed in swirling universes by using the techniques including the Poincar\'{e} section, fast Lyapunov exponent indicator, bifurcation diagram and basins of attraction. Comparing with the case of a Kerr black hole in Einstein's general relativity, we found that the swirling parameter $j$ makes the motion of particles more complex. We obtained the effects of the swirling parameter $j$ and the spin parameter $a$ on the dynamical behaviors of the motion of particles, and confirmed that there exists the chaotic motion for particles in the spacetimes of the Schwarzschild black hole and Kerr black hole immersed in swirling universes. We found that the chaotic region in Poincar\'{e} sections increases as the swirling parameter $j$ increases but decreases as the spin parameter $a$ increases, which shows that the parameters $j$ and $a$ have different effects on the motion of particles. Meanwhile, for the chaotic motions of orbits, 
we noted that as the swirling parameter $j$ (the spin parameter $a$) increases, both the lower bound of $a$ ($j$) and the range of the radial coordinate $r$ in the bifurcation diagram increase, which means that the presence of $j$ changes the range of $a$ where the chaotic motion appears for particles. Moreover, we observed clearly that, regardless of the spin parameter $a$, there exist some self-similar fractal fine structures in the basins boundaries of attractors, which indicates that there exists the chaotic motion for particles in the spacetime of a black hole immersed in swirling universes. Thus, the swirling parameter $j$ and the spin parameter $a$ yield the richer dynamical behavior of particles in the spacetime of a Kerr black hole immersed in swirling universes.

\begin{acknowledgments}

This work was supported by the National Natural Science Foundation of China (Grant Nos. 12275078, 12275079 and 12035005), National Key Research and Development Program of China (Grant No. 2020YFC2201400) and innovative research group of Hunan Province (2024JJ1006).

\end{acknowledgments}

\end{document}